\documentclass[aps,pra,onecolumn,nobibnotes,superscriptaddress,nofootinbib]{revtex4-2}
\usepackage{graphicx} 
\usepackage{caption,graphics,physics}
\usepackage{subcaption}
\usepackage{amsmath,xcolor,bm}
\usepackage{qcircuit}
\usepackage{mathtools,amsmath,amsthm,amsfonts,amssymb,amscd}
\usepackage{hyperref}
\usepackage[ruled,vlined]{algorithm2e}
\usepackage{algpseudocode}

\usepackage[capitalize]{cleveref}

\hypersetup{
    colorlinks=true,
    linkcolor=red,
    filecolor=magenta,      
    urlcolor=cyan,
}
\usepackage{geometry}

\Crefname{appsec}{Appendix}{Appendices}

\begin{document}
%
\title{Quantum medical image encoding and compression using Fourier-based methods}

\author{Taehee Ko}
\email{kthmomo@kias.re.kr}
\affiliation{Computational Sciences, Korea Institute for Advanced Study}

\author{Inho Lee}
\email{7tanyak@naver.com}
\affiliation{Department of Surgery, Seoul National University Bundang Hospital}

\author{Hyeong Won Yu}%
\email{hyeongwonyu@snu.ac.kr}
\affiliation{Department of Surgery, Seoul National University Bundang Hospital}

\begin{abstract}
      Quantum image processing (QIMP) has recently emerged as a promising field for modern image processing applications. In QIMP algorithms, encoding classical image informaiton into quantum circuit is important as the first step. However, most of existing encoding methods use gates almost twice the number of pixels in an image, and simulating even a modest sized image is computationally demanding. In this work, we propose a quantum image encoding method that effectively reduces gates than the number of pixels by a factor at least 4. We demonstrate our method for various 1024×1024 high-quality medical images captured during the Bilateral Axillo-Breast Approach (BABA) robotic thyroidectomy surgery. Additionally, two compression techniques are proposed to further reduce the number of gates as well as pre-processing time with negligible loss of image quality. We suggest our image encoding strategy as a valuable option for large scale medical imaging.

\end{abstract}
\maketitle

\section{Introduction}

With the advance of digital image technology, image processing has been routine computational disciplines in many applications such as computer vision \cite{nixon2019feature}, astronomy \cite{ramesh1986maximum}, and medical imaging \cite{sonka2013image}. Recently,  quantum image processing (QMIP) has drawn attention with potential image processing frameworks \cite{yan2024lessons,yan2020quantum} and the promise that quantum computing will bring a new computational paradigm inaccessible via classical computers \cite{wang2022review, nielsen2010quantum, lloyd1996universal}. In the last decade, numerous QIMP algorithms have been proposed, showing computational advantages than the classical counterparts  \cite{wang2015quantum,zhou2017global,liu2022quantum,yao2017quantum,chetia2021quantum,el2016quantum,zhou2017quantum}. Like these QIMP algorithms, developing image processing algorithms using classical and quantum resources together is expected as a new direction for improving modern medical imaging where the amount of image data provided daily is too large to be processed  (see \cite{yan2024review} and therein).

The key step in the QIMP algorithms is to encode classical image information into quantum circuit. Most of quantum image encoding methods proposed by far are based on two popular frameworks, FRQI \cite{le2011flexible} and NEQR \cite{zhang2013neqr}, while less has been explored with QPIE \cite{yao2017quantum}. Unfortunately, these encoding methods are not yet readily applicable even for modest sized images on the current quantum hardware but they also demand expensive computational cost for classical simulation. For example, with a FRQI-based encoding, the corresponding circuit may require at least two millions of gates including rotations and CNOTs for $1024\times 1024$ image, which is a typical size. This is due to the fact that the gate count required for FRQI is almost equal to the number of pixels in image  \cite{amankwah2022quantum}. This is the number far beyond the current capacitiy of quantum hardware. A NEQR-based encoding has an even worse gate complexity. It is the same for most of the previous quantum image encoding methods based on the framework of FRQI or NEQR \cite{khan2019improved,jiang2015quantum,sun2011multi,sang2017novel,amankwah2022quantum,haque2023advanced}. Therefore, it is important to find a quantum image encoding whose gate complexity is lower than the number of pixels for its feasibility for the future of QIMP and simulation purposes.  

\begin{figure}[htbp]

   \begin{subfigure}[b]{0.9\textwidth}
    \centering
    \includegraphics[width=\textwidth]{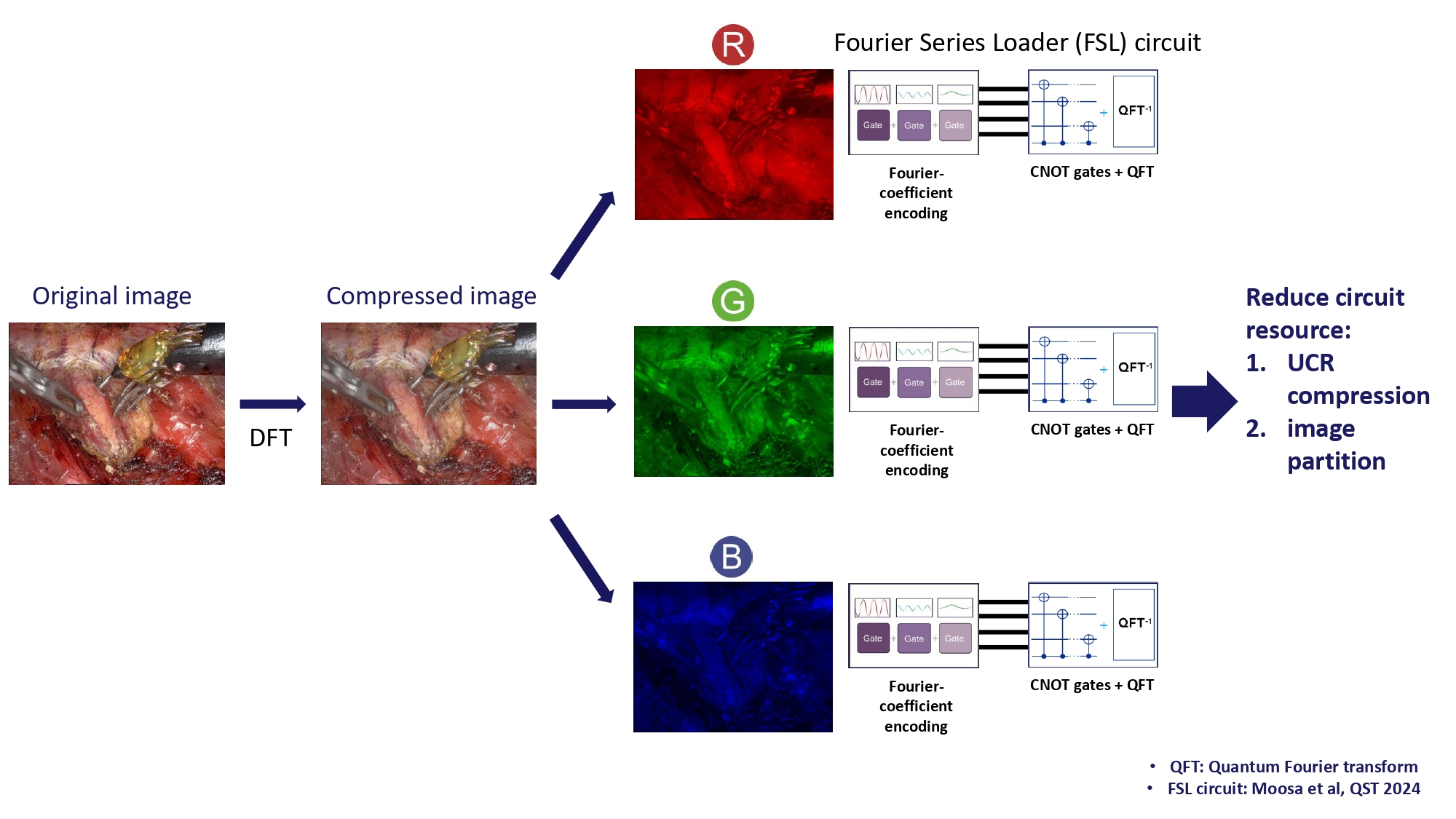}
    \end{subfigure}

    \caption{Workflow of the FAQPIE of RGB image and two options for reducing circuit resource}
    \label{fig:compare}
\end{figure}

\begin{figure}[htbp]

    \centering
    \begin{subfigure}[b]{0.3\textwidth}
    \centering
    \text{
\Qcircuit @C=0.9em @R=1.4em {
& \lstick{\ket{0^{n-m-1}}} & \qw{/} & \qw & \gate{X^{\otimes(n-m-1)}} \qw & \multigate{2}{QFT^\dagger} & \qw \\
& \lstick{\ket{0}} & \qw & \multigate{5}{\text{UCRs}} & \ctrl{-1} \qw & \ghost{QFT^\dagger} & \qw \\
& \lstick{\ket{0^m}} & \qw{/} & \ghost{\text{UCRs}} & \qw  & \ghost{QFT^\dagger} & \qw \\
& \\
& \lstick{\ket{0^{n-m-1}}} & \qw{/} & \qw & \gate{X^{\otimes(n-m-1)}} \qw & \multigate{2}{QFT^\dagger} & \qw\\
& \lstick{\ket{0}} & \qw & \ghost{\text{UCRs}} & \ctrl{-1} \qw & \ghost{QFT^\dagger} & \qw \\
& \lstick{\ket{0^m}} & \qw{/} & \ghost{\text{UCRs}} & \qw  & \ghost{QFT^\dagger} & \qw \\
}} 
    \caption{2d FSL circuit with UCRs}
    \end{subfigure}

    \begin{subfigure}[b]{0.6\textwidth}
    \centering
    \includegraphics[width=\textwidth]{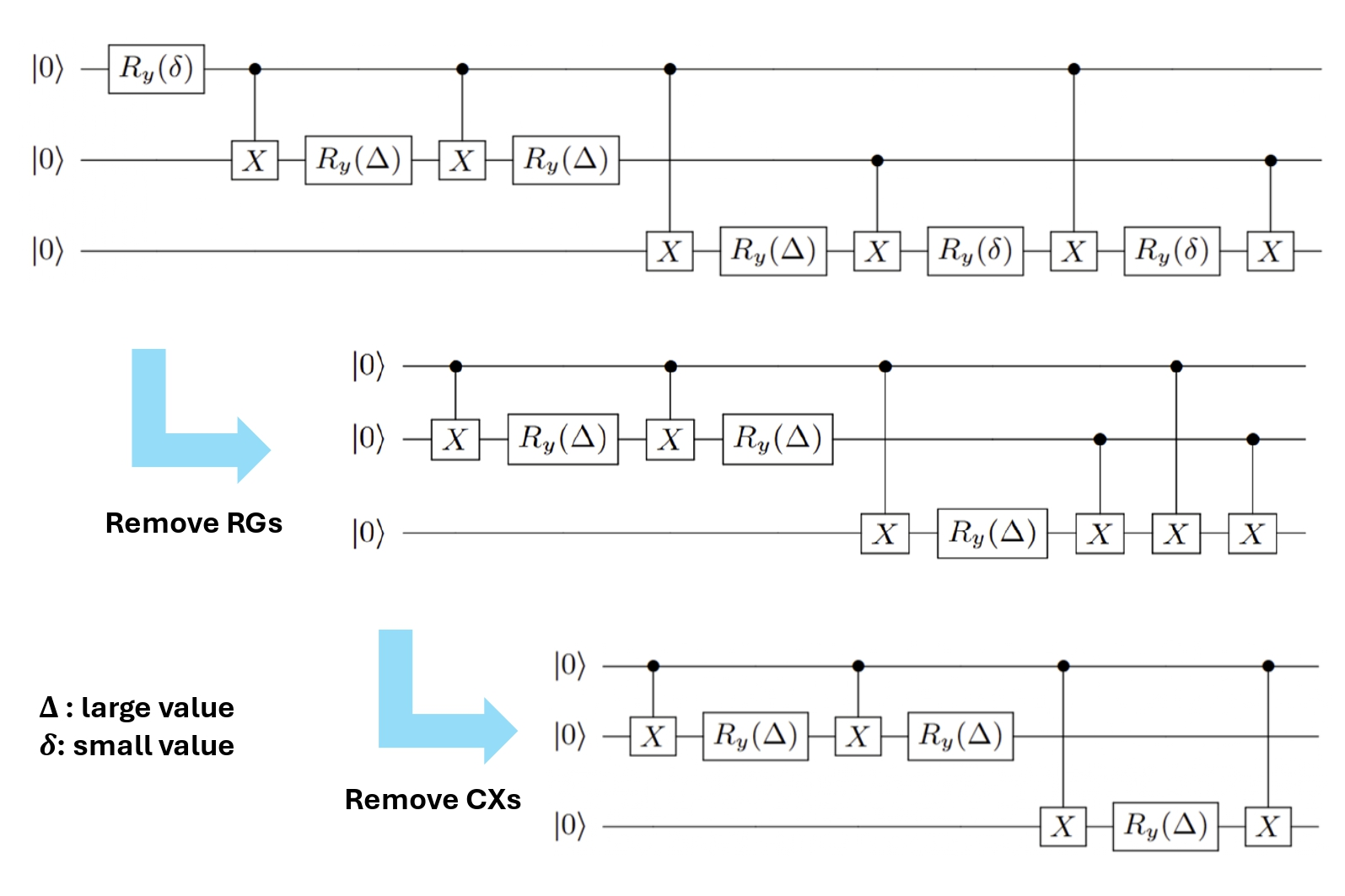}
    \caption{Compression of UCRs}
    \end{subfigure}
    
    \begin{subfigure}[b]{0.4\textwidth}
    \centering
    \includegraphics[width=\textwidth]{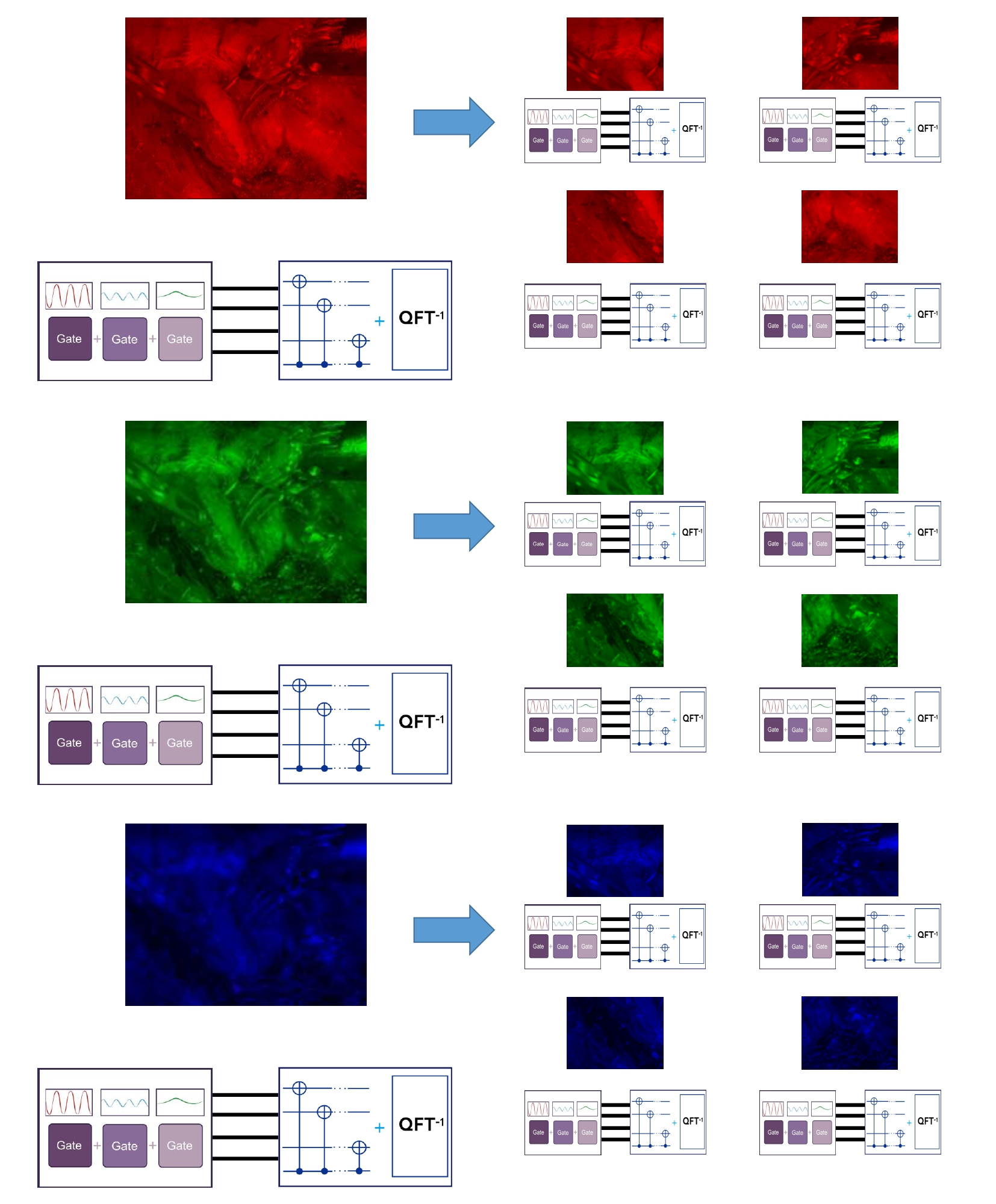}
    \caption{Image partition}
    \end{subfigure}
    \caption{(a) Illustration of 2d FSL circuit with uniformly-controlled rotations (UCRs) (b): Illustration of compression of a cascade of UCRs that are reformulated with rotation gates (RGs) and CNOT gates (CXs) (c): Image partition with reduced circuits for multiple FAQPIEs}
    \label{fig:compare1}
\end{figure}

In this paper, we propose a quantum image encoding method based on the discrete Fourier transform (DFT) using the FSL circuit \cite{moosa2023linear}.  We use the circuit as QPIE without any ancilla qubit for encoding images, embracing the idea of image compression based on the DFT \cite{hassanieh2012simple}. We call this Fourier-based approximate quantum probability image encoding (FAQPIE). 
Our experiment shows that the FAQPIE can achieve much lower gate counts than the aforementioned quantum image encoding methods. For example,  about 96\% reduction of gates is achieved compared to an improved version of FRQI \cite{amankwah2022quantum} for $1024\times 1024$ images. In addition, we use two compression strategies that further reduce gate count with little loss of image quality such as surgical details in medical image. One is based on image partition, inspired by applications using dicrete cosine transform (DCT) and DFT  \cite{haque2023advanced,gupta2012analysis,neethu2015enhancement}, and the other based on the compression of UCRs proposed in \cite{amankwah2022quantum}. Our experiment shows that a combination of these compression strategies reduces the maximal gate count up to about 80\% compared to the uncompressed case, without losing surgical details.  \cref{fig:compare} and \cref{fig:compare1} illustrate our  image encoding strategies, and \cref{table: precomparison} summarizes the proposed encoding method and others in terms of gate complexity and qubit number.

\begin{table}[htbp]

        \begin{tabular}{| c | c | c| c| c| c|}
        Encoding method  & QPIE \cite{mottonen2004transformation} & Improved FRQI \cite{amankwah2022quantum} & Improved NEQR \cite{amankwah2022quantum} &  FSL + FRQI \cite{moosa2023linear} & FAQPIE (this work)\\
  \hline Gate complexity & $\mathcal{O}(4^n)$ & $\mathcal{O}(4^n)$ & $\mathcal{O}(\ell 4^n)$ & $\mathcal{O}({\color{red}4^{m+2}}+n^2)$ & $\mathcal{O}({\color{blue}4^{m+2}}+n^2)$ \\
  \hline Total qubits & $2n$ & $2n+1$ & $2n+\ell$ & $2n+1$ & $2n$ \\
   \end{tabular}
    
     \caption{Summary of gate complexities and qubit count for encoding a $2^n\times 2^n$ image using different encoding methods. The QPIE is equivalent to preparing a generic quantum state. The $\ell$ denotes the bit depth in NEQR. The $m$ ($m\leq n-2$) is an adjustable parameter in \eqref{eq: imagrepm}, accounting for how many discrete Fourier coefficients of image are truncated. We note that the parameter $m$ (red) in \cite{moosa2023linear} depends on the smoothness of the trigonometic functions in FRQI, whereas in our method, it (blue) depends upon the compressibility of image based on DFT.}
     \label{table: precomparison}
\end{table}

\section{Method description}

\subsection{Approximate quantum probability image encoding via FSL circuit}

The quantum probability image encoding (QPIE) is an amplitude encoding that saves image pixels as the amplitudes of a quantum state. Specifically, the pixel values of image are encoded in probability amplitudes of a pure quantum state, and their positions correspond to the computational basis of the Hilbert space \cite{yao2017quantum}. 
For image $P=(P_{ij})_{N\times N}$ of $N\times N$ pixels with $N=2^n$, where $P_{kl}$ denotes the pixel value at position $(k,l)$, a QPIE is formulated as 
\begin{equation}\label{eq: P}
    \ket{P}=\frac{1}{\norm{P}_{F}}\sum_{k=0}^{2^n-1}\sum_{l=0}^{2^n-1}P_{kl}\ket{k}\otimes\ket{l}.
\end{equation}Implementing this QPIE in general requires an exponential scaling of gate complexity \cite{mottonen2004transformation}. Alternatively, we can prepare an approximation of the state \eqref{eq: P} using the FSL circuit \cite{moosa2023linear} with a lower gate complexity, assuming that the quality of image is reasonable after the DFT is applied. As will be discussed below, our image encoding method is different from the one in \cite[Appendix D]{moosa2023linear} in the sense that we consider discrete Fourier coefficients of image with QPIE while they consider those of trigonometric functions within FRQI. Therefore, the efficiency of our image encoding method is determined largely by image itself (e.g. its compressibility by DFT), while their method rely on the smoothness of the trigonometric functions at given pixel values. 

First, we can interpret the pixel as a function of two spatial variables $(k,l)$,   
\begin{equation}\label{eq: imagerep}
    P_{kl} = \sum_{x=0}^{2^n-1}\sum_{y=0}^{2^n-1}C_{xy}e^{i2\pi(\frac{kx}{2^n}+\frac{ly}{2^n})},\quad k,l\in[2^{n}-1]
\end{equation}where $C_{xy}$ denotes the Fourier coefficient with respect to frequencies $x,y$,
\begin{equation}\label{eq: Fourier coeff}
    C_{xy} = \sum_{k=0}^{2^n-1}\sum_{l=0}^{2^n-1}P_{kl}e^{-i2\pi(\frac{kx}{2^n}+\frac{ly}{2^n})}.
\end{equation}Second, we truncate relatively small Fourier coefficients in \eqref{eq: Fourier coeff}, which is  inspired by the idea of signal compression based on the discrete Fourier transform (DFT) \cite{hassanieh2012simple}. We then use FSL circuit to load the remaining large Fourier coefficients and define an approximation of QPIE \eqref{eq: P}, $\ket{P_m}$, to truncation order $m\leq n-2$
\begin{equation}\label{eq: imagrepm}
    \ket{P_m}=\frac{1}{\norm{P_m}_{F}}\sum_{k=0}^{2^n-1}\sum_{l=0}^{2^n-1}(P_m)_{kl}\ket{k}\otimes\ket{l},
\end{equation}where
\begin{equation} (P_m)_{kl}:=\sum_{x=0}^{2^m-1}\sum_{y=0}^{2^m-1}C_{xy}e^{i2\pi(\frac{kx}{2^n}+\frac{ly}{2^n})}.
\end{equation}We call this Fourier-based approximate quantum probability image encoding (FAQPIE). The upper bound of $m$ as $n-2$ is deduced from  the construction of FSL circuit.

\begin{figure}[htbp]

    \centering
   \begin{subfigure}[b]{0.19\textwidth}
    \centering
    \includegraphics[width=\textwidth]{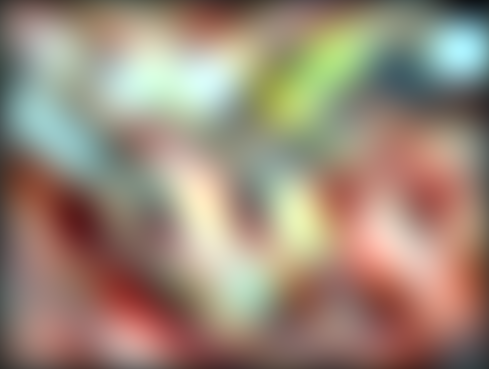}
    \end{subfigure}
    \centering
   \begin{subfigure}[b]{0.19\textwidth}
    \centering
    \includegraphics[width=\textwidth]{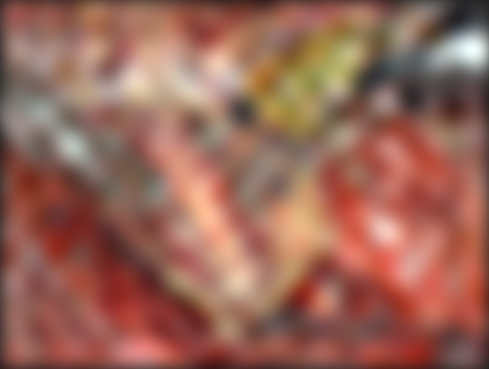}
    \end{subfigure}
    \centering
   \begin{subfigure}[b]{0.19\textwidth}
    \centering
    \includegraphics[width=\textwidth]{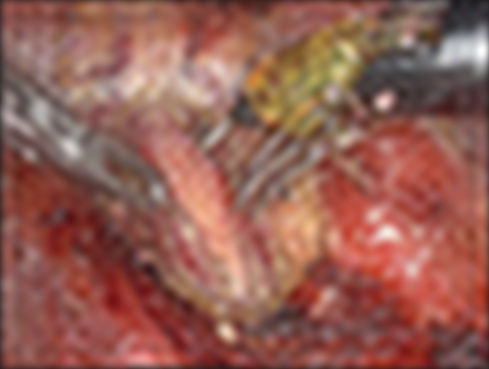}
    \end{subfigure}
    \centering
   \begin{subfigure}[b]{0.19\textwidth}
    \centering
    \includegraphics[width=\textwidth]{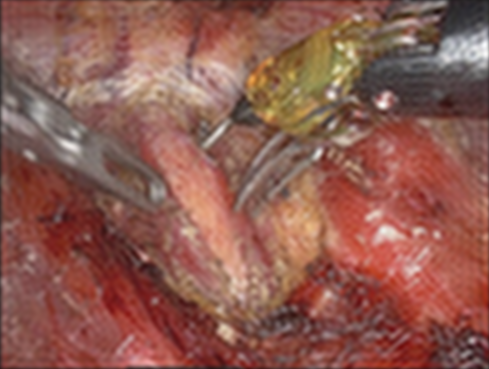}
    \end{subfigure}
    \centering
   \begin{subfigure}[b]{0.19\textwidth}
    \centering
    \includegraphics[width=\textwidth]{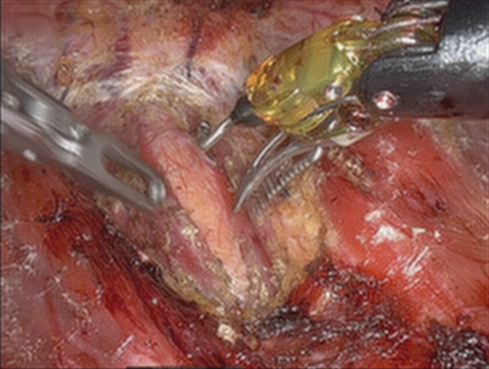}
    \end{subfigure}
    \caption{Illustration of the change of image quality of FAQPIE for a surgical image with increasing values of truncation order $m$ in \eqref{eq: imagrepm} from left to right ($m=3,4,5,6,7$).}
    \label{fig:mleast}
\end{figure}

The quality of FAQPIE depends on the truncation order $m$. A large value of $m$ ensures little loss of image information in FAQPIE. For example, in \cref{fig:mleast}, the size of the original image is $1024\times 1024$, which is the case $n=10$ in \eqref{eq: imagrepm}, and we apply the FAQPIE to the image with different values of $m$. We clearly see that the image quality becomes improved as $m$ icreases. As deduced from \cref{table: precomparison}, a large value of $m$ requires further quantum resource, and therefore there is a trade-off between the image quality of FAQPIE and the cost of constructing the corresponding circuit.  

\subsection{Compression strategies for FAQPIE}\label{sec: parallel}

The main component of FSL circuit is a unitary channel that prepares a quantum state whose amplitudes are the Fourier coefficients corresponding to low frequencies. For this, two types of channel are demonstrated in  \cite{moosa2023linear}: one based on UCRs \cite{mottonen2004transformation} and the other based on Schmidt decomposition \cite{krol2022efficient}. In this work, we focus on the UCR-based unitary channel for FAQPIE. In this section, we describe two relevant compression strategies.

\subsubsection{Compression of UCRs}

  In the UCR-based FSL circuit, the unitary channel consists of a cascade of UCRs that encodes discrete Fourier coefficients. Each UCR consists of alternating rotation gates and CNOT gates. The rotation gates act on the target qubit of the CNOT gates. Recently, \cite{amankwah2022quantum} introduces a compression method for this UCR circuit. The first step is to remove rotation gates whose parameters are sufficiently small. This leads to a good approximation of the original UCR. We use this idea to remove negligible rotation gates in the cascade of UCRs in the implementation of FAQPIE. The next step is to reduce CNOT gates by a parity check. Due to the elimination of rotation gates in the first step, some of consecutive CNOT gates can be cancelled out by using their commutativity and a parity check on the control qubits of the consecutive CNOT gates.

Notice that the reduction percentage of gates in UCRs with this compression technique depends on how many rotation gates are removed in the first step, that is, a thresholding rule that decides which parameters are meant to be small. In our numerical experiments, we followed a similar rule in \cite{amankwah2022quantum} that simply discard a certain percentage of the smallest parameters.

\subsubsection{Image partition}

The key idea of another compression technique is to partition an image into non-overlapping pieces and consider the respective FAQPIEs of these pieces. Basically, we construct a multiple of FAQPIE circuits for the image pieces. This would be useful in situations where implementing FAQPIE for the whole image is limited due to limited quantum hardware but encoding smaller image pieces into smaller circuits is feasible.

If an image of size $2^n\times 2^n$ is given, for some $n_0<n$, we can divide the image into the set of $2^{n_0}\times 2^{n_0}$ image pieces. Next we apply FFT to the pieces respectively and truncate their Fourier coefficients to some order $m$. After that, we obtain the set of FAQPIEs of the image pieces by defining FSL circuits, $\{\ket{P_{m,j}}\}_{j=0}^{4^{n-n_0}}$, whose classical sum equals the whole image \eqref{eq: imagrepm}, say,
\begin{equation}\label{eq: Parallel FSL}
    \ket{P_m} = \sum_{j=0}^{4^{n-n_0}}
 a_j\ket{j}\ket{P_{m,j}},\quad a_j = \frac{\norm{P_{m,j}}}{\norm{P_{m}}}.
\end{equation}Here $\ket{j}$ indicates the location of image piece $\ket{P_{m,j}}$. 

Specifically, we build multiple FSL circuits for the FAQPIEs of image pieces, and consider the maximum of the numbers of gates, qubits, and the total pre-processing time as measures when multiple image encoding circuits are required,
\begin{equation}\label{eq: maximal criterion}
    \max_{j}\{\text{\# of gates for }\ket{P_{m,j}}\},\; \max_j\{\text{\# of qubits for }\ket{P_{m,j}}\}\;, \sum_j\left(\text{Pre-processing time for }\ket{P_{m,j}}\right).
\end{equation}
As we will show in numerical experiments, the idea of encoding image pieces allows circuit compression in terms of maximal circuit complexity and total pre-processing time.  The first advantage of image partition \eqref{eq: Parallel FSL} is that the number of qubits required decreases by $2(n-n_0)$, since the image pieces are to be encoded with $2n_0$ qubits. Secondly, as we will demonstrate in our experiment, the truncation order $m$ can be reduced without largely degrading image quality. This allows for a smaller truncation order $m_0<m$ for $\ket{P_m}$ in \eqref{eq: Parallel FSL}, and consequently, the total pre-procesing time is significantly reduced. Therefore, the image paritition approach \eqref{eq: Parallel FSL} leads to more efficient circuit constructions, compared to the case of the  whole image encoding \eqref{eq: imagrepm}.

\section{Numerical experiments}

In this section, we demonstrate the proposed encoding strategies with several medical images. By incorporating the strategies into the code for FSL circuit \cite{moosa2023linear}  without introducing ancilla qubit, we conduct numerical simulations based on the function "Statevector" in Qiskit 1.2.4, and compare the simulation results. As mentioned earlier, we use FSL circuit with a unitary channel based on the UCRs \cite{mottonen2004transformation}.

The medical images considered are obtained from a video of a surgery called Bilateral Axillo-Breast Approach (BABA) robotic thyroidectomy, approved by the Research Ethics Committee of Seoul National University Bundang Hospital (B-2504-969-701). This is a thyroid surgery method that is representative of remote access surgery. Clinically, BABA surgery is well known for its superior cosmetic effects and lower complications than conventional thyroidectomy. This surgical method creates surgical incisions in both axillae and both areolas and uses the da Vinci robot to perform the surgery. The surgical video is recorded during the surgery, and the high-resolution images captured are analyzed to develop meaningful clinical studies. Therefore, encoding a high quality of image into a quantum circuit may be significantly important for the future of clinical studies.

\begin{figure}[htbp]

    \centering
   \begin{subfigure}[b]{0.22\textwidth}
    \centering
    \includegraphics[width=\textwidth]{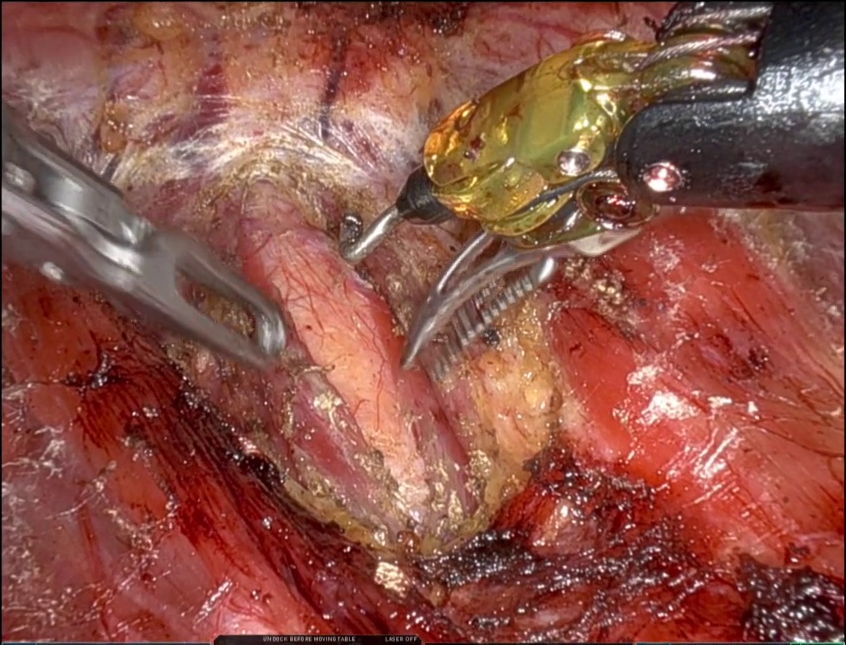}
    \caption{Original Image}
    \end{subfigure}
    \centering
   \begin{subfigure}[b]{0.22\textwidth}
    \centering
    \includegraphics[width=\textwidth]{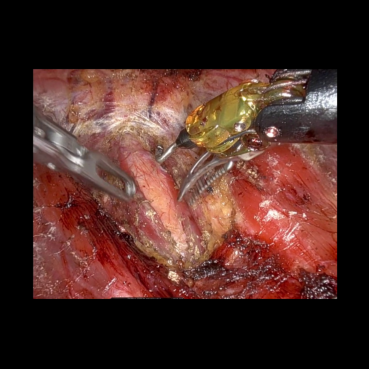}
    \caption{Original Image with zero padding}
    \end{subfigure}
    
    \centering
   \begin{subfigure}[b]{0.22\textwidth}
    \centering
    \includegraphics[width=\textwidth]{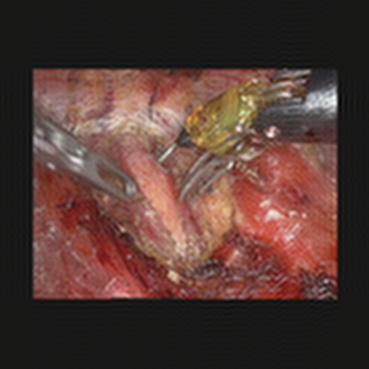}
    \caption{FAQPIE}
    \end{subfigure}
    \centering
   \begin{subfigure}[b]{0.22\textwidth}
    \centering
    \includegraphics[width=\textwidth]{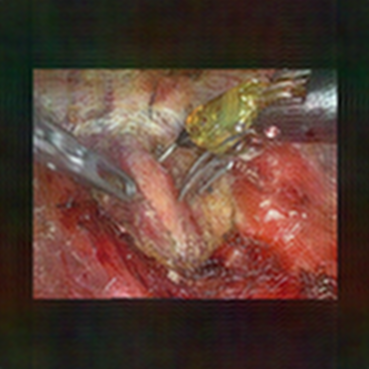}
    \caption{FAQPIE (CUCR)}
    \end{subfigure}
    \centering
   \begin{subfigure}[b]{0.22\textwidth}
    \centering
    \includegraphics[width=\textwidth]{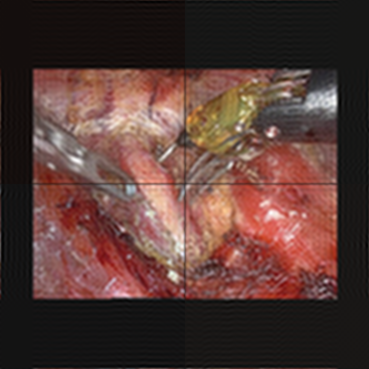}
    \caption{FAQPIE (IP)}
    \end{subfigure}
    \centering
   \begin{subfigure}[b]{0.22\textwidth}
    \centering
    \includegraphics[width=\textwidth]{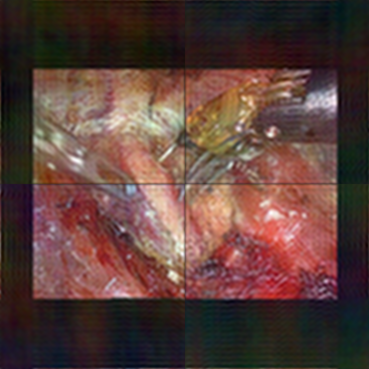}
    \caption{FAQPIE (CUCR + IP)}
    \end{subfigure}
    \caption{We zero pad the original medical images (a) to one of size $1024\times 1024$ (b). Then, we encode the zero-padded surgical image using FAQPIE for the simulation (c). The figures (d), (e), and (f) are obtained with one of the compression techniques or both in \cref{sec: parallel}. For short, we call compression of UCRs CUCR and image partition IP. }
    \label{fig:frames}
\end{figure}

The image \cref{fig:frames}(a) has a resolution of $636\times 842$ pixels as cropped from a $1280\times 720$ image, which shows both human organ and surgical tools. We zero pad the image to one with $1024\times 1024$ pixels as shown in \cref{fig:frames}(b). Since the new image has $1024\times 1024$ pixels, the required number of qubits for \cref{fig:frames}(c),(d) is 20. On the other hand, for the images \cref{fig:frames}(e),(f), which correspond to the case of blockwise image encoding strategy, we divide the new image into $512\times 512$ four images. Thus, 18 qubits are required for the simulated images in \cref{fig:frames}(e),(f).

Since the image is colored, we apply the encoding strategies three times by dividing a 3d-array of pixels involving RGB colors into three 2d arrays. Thus, for \cref{fig:frames}(c),(d), which is the case of encoding the whole image, 3 circuits are used, while 12 circuits are used for \cref{fig:frames}(e),(f) since the image are partitioned into four blocks.

As seen in \cref{fig:mleast}, we should select a reasonable truncation order $m$ such that a given encoding strategy preserves surgical details of medical image. To validate the choice of parameter $m$, we define a metric to measure the qualty of divided image encoding in \eqref{eq: Parallel FSL} as follows,
\begin{equation}\label{eq: fidelity}
\text{Fidelity}=\frac{\sum_{j=0}^J\abs{\bra{P_{m,j}}\ket{P_j}}^2}{J}.
\end{equation}This quantity means the average of fidelities of the FAQPIEs for approximate image pieces, $\ket{P_{m,j}}$'s in \eqref{eq: Parallel FSL}, with the corresponding pieces $\ket{P_j}$'s of the original image $\ket{P}$ in \eqref{eq: imagerep}. The averaged fidelity \eqref{eq: fidelity} equals the usual fidelity \cite{moosa2023linear} when $J=0$, 
\begin{equation}
    \abs{\bra{P_m}\ket{P}}^2,
\end{equation}which means no image partition.
With this figure of merit, \cref{table: comparison1} shows  high fidelity results for all simulated images, which means that the values of $m$ in our simulations are set reasonably.

We note if the exact encoding of $1024\times 1024$ medical image (i.e. QPIE) is desired using the UCR gates, the required numbers of single-qubit gates (e.g. Ry, Rz), and CNOT gates are $2097148=2^{21}-4$, which can be deduced from the function "cascade\_UCRs" in the code \cite{moosa2023linear}. Compared to this large number, the proposed encoding strategies use at most $65528=2^{16}-8$ gates, which is roughly 98\% reduction of gates, as in the first case in \cref{table: comparison1}.

\cref{table: comparison1} compares the uncompressed case of FAQPIE to the others with the compressed strategies: compression of UCRs (CUCR) and image partition (IP).  As observed, the compressed strategies are very effective for the image. In particular, when the IP is applied, significant reductions of gates per circuit are obtained  without losing surgical details, and even further reductions of gates are observed when the CUCR is further applied, as shown in \cref{fig:frames}(e),(f). Another advantage of the IP is that 
the pre-processing time can be reduced significantly. We observe that the elapsed time is reduced by about a factor of 4. The reason is that a lower truncation order still guarantees reasonable quality of image, thereby yielding a significant reduction of the time for constructing UCRs in circuits. Lastly, looking at the pre-processing times in \cref{table: comparison1},
we note that the time for the compression threshold is negligible compared to the time spent constructing the UCRs.

\begin{figure}[htbp]

    \centering
   \begin{subfigure}[b]{0.3\textwidth}
    \centering
    \includegraphics[width=\textwidth]{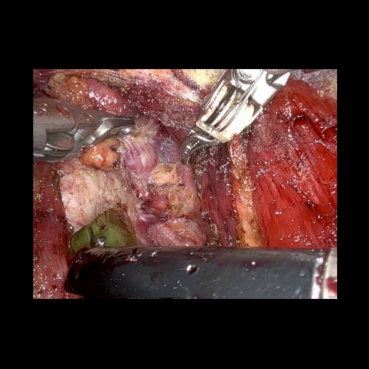}
    \end{subfigure}
    \begin{subfigure}[b]{0.3\textwidth}
    \centering
    \includegraphics[width=\textwidth]{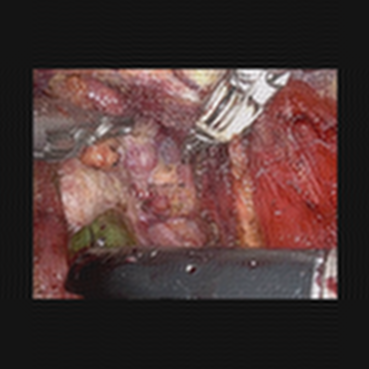}
    \end{subfigure}
   \begin{subfigure}[b]{0.3\textwidth}
    \centering
    \includegraphics[width=\textwidth]{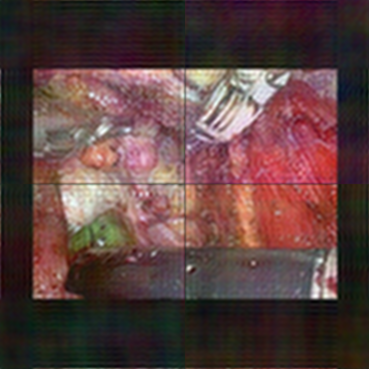}
    \end{subfigure}

    \centering
   \begin{subfigure}[b]{0.3\textwidth}
    \centering
    \includegraphics[width=\textwidth]{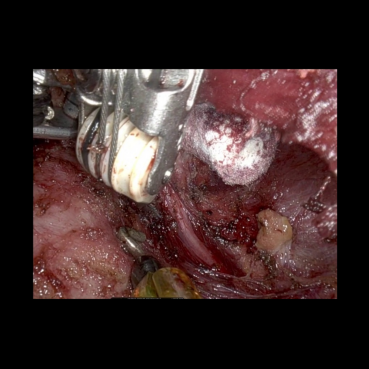}
    \end{subfigure}
    \begin{subfigure}[b]{0.3\textwidth}
    \centering
    \includegraphics[width=\textwidth]{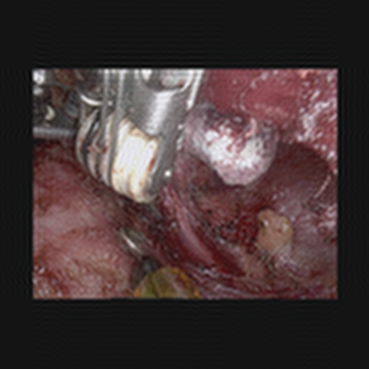}
    \end{subfigure}
   \begin{subfigure}[b]{0.3\textwidth}
    \centering
    \includegraphics[width=\textwidth]{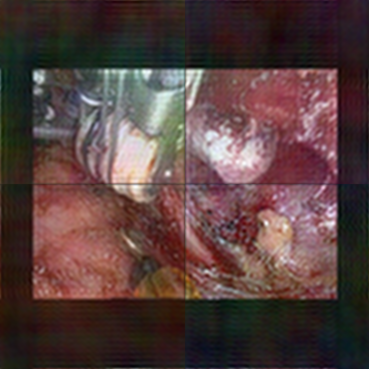}
    \end{subfigure}

    \centering
   \begin{subfigure}[b]{0.3\textwidth}
    \centering
    \includegraphics[width=\textwidth]{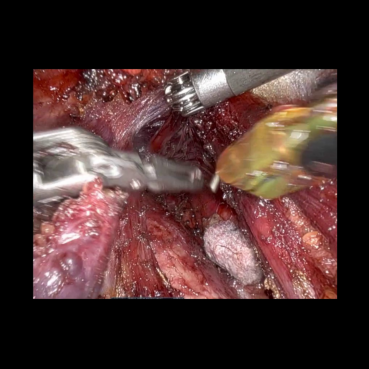}
    \end{subfigure}
    \begin{subfigure}[b]{0.3\textwidth}
    \centering
    \includegraphics[width=\textwidth]{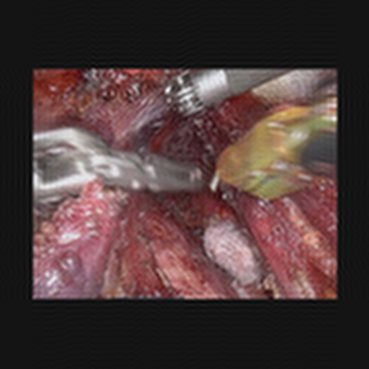}
    \end{subfigure}
   \begin{subfigure}[b]{0.3\textwidth}
    \centering
    \includegraphics[width=\textwidth]{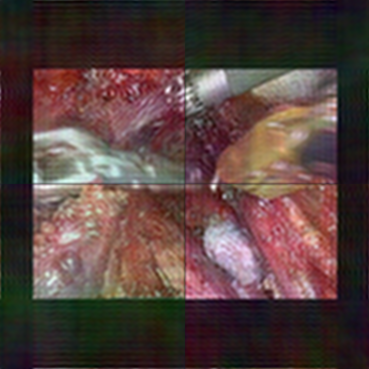}
    \end{subfigure}

    \centering
   \begin{subfigure}[b]{0.3\textwidth}
    \centering
    \includegraphics[width=\textwidth]{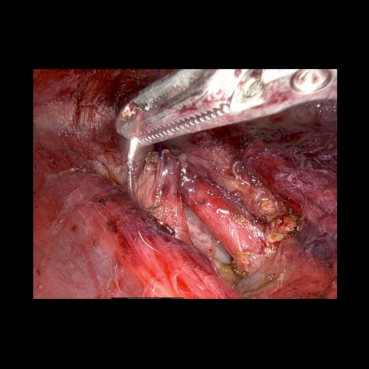}
    \end{subfigure}
    \begin{subfigure}[b]{0.3\textwidth}
    \centering
    \includegraphics[width=\textwidth]{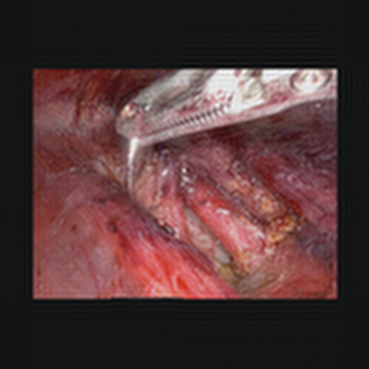}
    \end{subfigure}
   \begin{subfigure}[b]{0.3\textwidth}
    \centering
    \includegraphics[width=\textwidth]{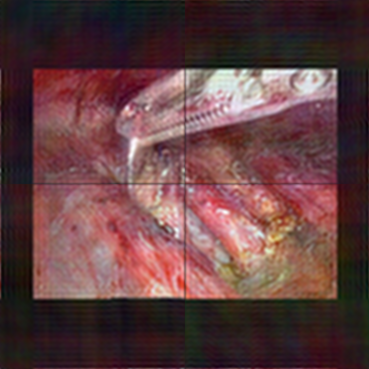}
    \end{subfigure}
    \caption{Comparison of original medical images (left), the corresponding reconstructed images via FAQPIE with no compression technique (middle) and the two compression techniques applied as in \cref{fig:frames} (right). }
    \label{fig:frame2s}
\end{figure}

We performed similar numerical simulations for more images taken at different frames. \cref{fig:frame2s} compares the original zero-padded images, the corresponding reconstructed images using FAQPIE with and without the combination of two compression techniques with the same parameter setting as in \cref{fig:frames}. A similar observation is made that encoding the images with the two compression techniques results in more efficient FAQPIE with reasonable surgical details and lowered gate counts, as summarized \cref{table: comparison}.

\begin{table}[htbp]

        \begin{tabular}{| c | c | c | c | c |}
        \hline 
        Encoding strategy   & FAQPIE & FAQPIE (CUCR) & FAQPIE (IP) & FAQPIE (CUCR + IP)\\
 \hline Qubits  & 20
 & 20 & 18 & 18\\
 \hline Circuit count  & 3 & 3 & 12 & 12  \\
 \hline truncation order ($m$)  & 6
 & 6 & 5 & 5\\
  \hline Fidelity(R)  & 0.9931 & 0.9753 & 0.9815 & 0.9392  \\
  \hline Fidelity(G)  & 0.9836 & 0.9553 & 0.9605 & 0.9296  \\
  \hline Fidelity(B)  & 0.9789 & 0.9433 & 0.9495 & 0.9012  \\
 \hline Pre-processing time (sec)   & 1355.69 & 1381.90 & 300.71  & 313.58 \\
 \hline Ry+Rz (maximal)  & 32764
 & 22948 & 8188 & 5741 \\
  \hline CNOT (maximal)   & 32764 & 29846  & 8188  & 7478 \\
 \hline Ry+Rz reduction  & 0\% & 29.95\% & 75.00\% & 82.47\%  \\
 \hline CNOT reduction  & 0\% & 8.90\% & 75.00\% & 77.17\%  \\
 \hline
   \end{tabular}
    
     \caption{Comparison of several encoding strategies for the task of reconstructing the original image of images in \cref{fig:mleast}. For each simulation, a reasonable truncation order $m$ is selected to capture details of medical image. The fidelity is defined in \eqref{eq: fidelity}. The number of circuits required to encode the whole image is also presented. For the circuits, the maximal number of qubits is shown. The pre-processing time is the total time spent finding parameters in UCRs \cite{mottonen2004transformation}. Among the circuits, we count the maximal number of single-qubit gates and that of CNOT gates before a cascade of CNOT gates and the inverse QFT in the FSL circuit. Compared to the gate numbers in FAQPIE, the corresponding gate reduction is estimated. }
     \label{table: comparison1}
\end{table}

\begin{table}[htbp]

        \begin{tabular}{| c | c | c | c | c |}
        \hline 
        Image   & 1 & 2 & 3 & 4\\
\hline Ry+Rz (maximal)  & 5742
 & 5742 & 5742 & 5741 \\
  \hline CNOT (maximal)   & 7484 & 7472  & 7506  & 7488 \\
 \hline Ry+Rz reduction  & 82.47\% & 82.47\% & 82.47\% & 82.47\%  \\
 \hline CNOT reduction  & 77.15\% & 77.19\% & 77.09\% & 77.14\%  \\
 \hline
   \end{tabular}
    
     \caption{Gates and reduction percentages for the simulation results in \cref{fig:frame2s}, which are indicated similar to the ones in \cref{table: comparison1}. }
     \label{table: comparison}
\end{table}

\section{Conclusion}
Inspired by  the idea of image compression using the discrete Fourier transform (DFT), we considered the use of Fourier series loader circuit \cite{moosa2023linear} as an approximate quantum probability image encoding. This Fourier-based approximate quantum image probability encoding (FAQPIE) requires $\mathcal{O}(4^{m+2}+n^2)$ gate complexity for 
a $2^n\times 2^n$ image where the parameter $m$ is determined by the efficiency of quantum circuit and the quality of image. In fact, FAQPIE becomes more efficient for images that are effectively compressible based on DFT. We conducted numerical simulations for various surgical images that involve complex medical features. Of more practical interest, two compression strategies are incorporated to further reduce the number of gates. By extending a circuit compression strategy for uniformly-controlled rotation gates \cite{amankwah2022quantum}, we are able to reduce the gate complexity of a cascade of uniformly-controlled rotation gates within the proposed quantum image encoding strategy.  Another compression strategy is based on image partition, which allows for further truncation of discrete Fourier coefficients and significantly reduces maximal gate complexity and classical pre-processing time. Combining the two compression strategies, we were able to achieve about 80\% reduction of maximal gate complexity compared to the uncompressed strategy, without sacrificing details of surgical images. The FAQPIE may be a valuable option for quantum image processing such as edge detection. For example, our encoding method may complement the quantum hadamard edge detection algorithm (QHED) \cite{yao2017quantum} where no general circuit is provided for the quantum probability image encoding. In addition, it would be interesting to study whether an  approach using the notion of matrix-product-state (MPS) \cite{jobst2024efficient,dilip2022data} togerther with our compression techniques can further reduce circuit complexity for various medical images.  








\section{Acknowledgment}
We used resources of the Center for Advanced Computation at Korea Institute for Advanced Study and the National Energy Research Scientific Computing Center (NERSC), a U.S. Department of Energy Office of Science User Facility operated under Contract No. DE-AC02-05CH11231. T.K. is supported by a KIAS Individual Grant (No. CG096001) at Korea Institute for Advanced Study.

\section{Author contributions}
The project was conceived by T.K. and H.W.Y.  Numerical implementations were performed by T.K. The manuscript was written by T.K., I.L., and H.W.Y.

\bibliographystyle{plain}
\bibliography{qc}

\begin{thebibliography}{10}

\bibitem{amankwah2022quantum}
Mercy~G Amankwah, Daan Camps, E~Wes Bethel, Roel Van~Beeumen, and Talita Perciano.
\newblock Quantum pixel representations and compression for n-dimensional images.
\newblock {\em Scientific reports}, 12(1):7712, 2022.

\bibitem{chetia2021quantum}
Rajib Chetia, SMB Boruah, and PP~Sahu.
\newblock Quantum image edge detection using improved sobel mask based on neqr.
\newblock {\em Quantum information processing}, 20(1):21, 2021.

\bibitem{dilip2022data}
Rohit Dilip, Yu-Jie Liu, Adam Smith, and Frank Pollmann.
\newblock Data compression for quantum machine learning.
\newblock {\em Physical Review Research}, 4(4):043007, 2022.

\bibitem{el2016quantum}
Abdelilah El~Amraoui, Lhoussaine Masmoudi, Hamid Ez-Zahraouy, and Youssef El~Amraoui.
\newblock Quantum edge detection based on shannon entropy for medical images.
\newblock In {\em 2016 IEEE/ACS 13th International Conference of Computer Systems and Applications (AICCSA)}, pages 1--6. IEEE, 2016.

\bibitem{gupta2012analysis}
Maneesha Gupta and Amit~Kumar Garg.
\newblock Analysis of image compression algorithm using dct.
\newblock {\em International Journal of Engineering Research and Applications (IJERA)}, 2(1):515--521, 2012.

\bibitem{haque2023advanced}
Md~Ershadul Haque, Manoranjan Paul, Anwaar Ulhaq, and Tanmoy Debnath.
\newblock Advanced quantum image representation and compression using a dct-efrqi approach.
\newblock {\em Scientific Reports}, 13(1):4129, 2023.

\bibitem{hassanieh2012simple}
Haitham Hassanieh, Piotr Indyk, Dina Katabi, and Eric Price.
\newblock Simple and practical algorithm for sparse fourier transform.
\newblock In {\em Proceedings of the twenty-third annual ACM-SIAM symposium on Discrete Algorithms}, pages 1183--1194. SIAM, 2012.

\bibitem{jiang2015quantum}
Nan Jiang and Luo Wang.
\newblock Quantum image scaling using nearest neighbor interpolation.
\newblock {\em Quantum Information Processing}, 14:1559--1571, 2015.

\bibitem{jobst2024efficient}
Bernhard Jobst, Kevin Shen, Carlos~A Riofr{\'\i}o, Elvira Shishenina, and Frank Pollmann.
\newblock Efficient mps representations and quantum circuits from the fourier modes of classical image data.
\newblock {\em Quantum}, 8:1544, 2024.

\bibitem{khan2019improved}
Rabia~Amin Khan.
\newblock An improved flexible representation of quantum images.
\newblock {\em Quantum Information Processing}, 18:1--19, 2019.

\bibitem{krol2022efficient}
Anna~M Krol, Aritra Sarkar, Imran Ashraf, Zaid Al-Ars, and Koen Bertels.
\newblock Efficient decomposition of unitary matrices in quantum circuit compilers.
\newblock {\em Applied Sciences}, 12(2):759, 2022.

\bibitem{le2011flexible}
Phuc~Q Le, Fangyan Dong, and Kaoru Hirota.
\newblock A flexible representation of quantum images for polynomial preparation, image compression, and processing operations.
\newblock {\em Quantum Information Processing}, 10:63--84, 2011.

\bibitem{liu2022quantum}
Wenjie Liu and Lu~Wang.
\newblock Quantum image edge detection based on eight-direction sobel operator for neqr.
\newblock {\em Quantum information processing}, 21(5):190, 2022.

\bibitem{lloyd1996universal}
Seth Lloyd.
\newblock Universal quantum simulators.
\newblock {\em Science}, 273(5278):1073--1078, 1996.

\bibitem{moosa2023linear}
Mudassir Moosa, Thomas~W Watts, Yiyou Chen, Abhijat Sarma, and Peter~L McMahon.
\newblock Linear-depth quantum circuits for loading fourier approximations of arbitrary functions.
\newblock {\em Quantum Science and Technology}, 9(1):015002, 2023.

\bibitem{mottonen2004transformation}
Mikko Mottonen, Juha~J Vartiainen, Ville Bergholm, and Martti~M Salomaa.
\newblock Transformation of quantum states using uniformly controlled rotations.
\newblock {\em arXiv preprint quant-ph/0407010}, 2004.

\bibitem{neethu2015enhancement}
S~Neethu, S~Sreelakshmi, and Deepa Sankar.
\newblock Enhancement of fingerprint using fft$\times$| fft| n filter.
\newblock {\em Procedia Computer Science}, 46:1561--1568, 2015.

\bibitem{nielsen2010quantum}
Michael~A Nielsen and Isaac~L Chuang.
\newblock {\em Quantum computation and quantum information}.
\newblock Cambridge university press, 2010.

\bibitem{nixon2019feature}
Mark Nixon and Alberto Aguado.
\newblock {\em Feature extraction and image processing for computer vision}.
\newblock Academic press, 2019.

\bibitem{ramesh1986maximum}
Narayan Ramesh and Rajaram Nityananda.
\newblock Maximum entropy image restoration in astronomy.
\newblock {\em Annual review of astronomy and astrophysics}, 24:127--170, 1986.

\bibitem{sang2017novel}
Jianzhi Sang, Shen Wang, and Qiong Li.
\newblock A novel quantum representation of color digital images.
\newblock {\em Quantum Information Processing}, 16:1--14, 2017.

\bibitem{sonka2013image}
Milan Sonka, Vaclav Hlavac, and Roger Boyle.
\newblock {\em Image processing, analysis and machine vision}.
\newblock Springer, 2013.

\bibitem{sun2011multi}
Bo~Sun, Phuc~Q Le, Abdullah~M Iliyasu, Fei Yan, J~Adrian Garcia, Fangyan Dong, and Kaoru Hirota.
\newblock A multi-channel representation for images on quantum computers using the rgb$\alpha$ color space.
\newblock In {\em 2011 IEEE 7th International Symposium on Intelligent Signal Processing}, pages 1--6. IEEE, 2011.

\bibitem{wang2015quantum}
Jian Wang, Nan Jiang, and Luo Wang.
\newblock Quantum image translation.
\newblock {\em Quantum Information Processing}, 14(5):1589--1604, 2015.

\bibitem{wang2022review}
Zhaobin Wang, Minzhe Xu, and Yaonan Zhang.
\newblock Review of quantum image processing.
\newblock {\em Archives of Computational Methods in Engineering}, 29(2):737--761, 2022.

\bibitem{yan2024review}
Fei Yan, Hesheng Huang, Witold Pedrycz, and Kaoru Hirota.
\newblock Review of medical image processing using quantum-enabled algorithms.
\newblock {\em Artificial Intelligence Review}, 57(11):300, 2024.

\bibitem{yan2020quantum}
Fei Yan and Salvador~E Venegas-Andraca.
\newblock {\em Quantum image processing}.
\newblock Springer Nature, 2020.

\bibitem{yan2024lessons}
Fei Yan and Salvador~E Venegas-Andraca.
\newblock Lessons from twenty years of quantum image processing.
\newblock {\em ACM Transactions on Quantum Computing}, 2024.

\bibitem{yao2017quantum}
Xi-Wei Yao, Hengyan Wang, Zeyang Liao, Ming-Cheng Chen, Jian Pan, Jun Li, Kechao Zhang, Xingcheng Lin, Zhehui Wang, Zhihuang Luo, et~al.
\newblock Quantum image processing and its application to edge detection: theory and experiment.
\newblock {\em Physical Review X}, 7(3):031041, 2017.

\bibitem{zhang2013neqr}
Yi~Zhang, Kai Lu, Yinghui Gao, and Mo~Wang.
\newblock Neqr: a novel enhanced quantum representation of digital images.
\newblock {\em Quantum information processing}, 12:2833--2860, 2013.

\bibitem{zhou2017quantum}
Ri-Gui Zhou, Xingao Liu, and Jia Luo.
\newblock Quantum circuit realization of the bilinear interpolation method for gqir.
\newblock {\em International Journal of Theoretical Physics}, 56:2966--2980, 2017.

\bibitem{zhou2017global}
Ri-Gui Zhou, Canyun Tan, and Hou Ian.
\newblock Global and local translation designs of quantum image based on frqi.
\newblock {\em International Journal of Theoretical Physics}, 56:1382--1398, 2017.

\end{thebibliography}

\end{document}